\documentstyle[12pt]{article}
\jot = 1.5ex\catcode`\@=11\renewcommand{\thefootnote}{\fnsymbol{footnote}}

\def\be{\begin{equation}}
\def\ee{\end{equation}}
\def\l{\label}
\def\F{{\cal F}}

\def\ldl{\Lambda\partial_\Lambda}

\begin{document}\begin{titlepage}

\rightline{hep-th/9712039}
\rightline{DFPD97/TH/55}

\vspace{1.cm}

\centerline{\large{\bf Beta Function, C--Theorem}}

\vspace{.5cm}

\centerline{\large{\bf and WDVV Equations in 4D N=2 SYM}}

\vspace{1.5cm}

{\centerline{\sc Gaetano BERTOLDI and Marco MATONE}}

\vspace{0.3cm}

\centerline{\it Department of Physics ``G. Galilei'' and INFN, Padova, 
Italy}
\centerline{bertoldi@padova.infn.it  matone@padova.infn.it}

\vspace{2cm}

\centerline{\bf ABSTRACT}

\vspace{.6cm}
\noindent
We show that the exact $beta$--function of 4D $N=2$ SYM 
plays the role 
of the metric whose inverse satisfies the 
WDVV--like equations $\F_{ikl}\beta^{lm} \F_{mnj}=
\F_{jkl}\beta^{lm}\F_{mni}$. The conjecture that the WDVV--like equations 
are equivalent to the identity involving the $u$--modulus and the 
prepotential $\F$,
seen as a superconformal anomaly, sheds light 
on the recently considered c--theorem for the $N=2$ SYM field theories.

\vspace{0.6cm}
\noindent

\end{titlepage}

\newpage
\setcounter{footnote}{0}
\renewcommand{\thefootnote}{\arabic{footnote}}

The Seiberg--Witten results about $N=2$ SUSY Yang--Mills 
\cite{SW1} have been recently rederived, in the $SU(2)$ case, in \cite{BMT}.
The approach, based on uniformization theory, uses reflection symmetry
of quantum vacua, 
asymptotics analysis and the
identity \cite{mat}
\begin{equation}
u=\pi i ({\cal F}-a\partial_a{\cal F}/2).
\label{81}\end{equation}
This identity,
first checked up to 
two--instanton in \cite{FUTR}, has been proved to 
any order in the instanton expansion in \cite{DOKHMA}. Furthermore,
Eq.(\ref{81}) has been
obtained as an anomalous superconformal
Ward identity in \cite{HW}.

As a consequence of the  Seiberg--Witten results, it has been possible
to derive the explicit expression for the $beta$--function 
\cite{Loro}\cite{Noi}\cite{Noi2}\cite{DKP} which has been 
recently reconsidered in \cite{Essi1}\cite{Essi2}\cite{Essi3}. 

The above results suggested looking for 
the analogue of the Zamolodchikov c--theorem
\cite{zam} in the framework of $4D$ N=2 SYM.
In particular, 
very recently, 
it has been shown in \cite{GIMA} that the results in \cite{Essi2}
can be understood from the c--theorem point of view
(see \cite{varii} for related aspects). 
Furthermore, it has been shown that for the $SU(n)$ case
there is a Lyapunov function which is naturally determined and
related to the {\it
classical} discriminant of the Seiberg--Witten curve.
It has been also observed that the c--theorem point of view actually fits
with the fact that,
according to \cite{HW}, Eq.(\ref{81}) means that 
$u$ is proportional to the (super)conformal anomaly.

In this paper, we first shortly consider the role of the $beta$--function
in the framework of the WDVV--like equations which have been introduced in 
\cite{Noi2} for the $SU(3)$ case. While for $SU(3)$ the $beta$--function
satisfies a basic equation, derived from the reduced
Picard--Fuchs equations, the identification of the inverse of the 
$beta$--function
with the WDVV metric results in the identity
\be
\F_{ikl}\beta^{lm} \F_{mnj}=
\F_{jkl}\beta^{lm}\F_{mni}.
\l{bw}\ee
This naturally suggests
considering this equation for the groups $SU(n)$ $n>3$.
It turns out that in this case (\ref{bw}) is no longer an identity and actually
is a consequence of the WDVV--like equations derived in \cite{MAMIMO}
(see \cite{Carroll} for related aspects).

The appearance of the $beta$--function in (\ref{bw}), and the way we derive
it, suggests considering it as related to the superconformal anomaly. 
This would be in agreement
with the general setting considered in \cite{GIMA} and the 
results in \cite{mat}\cite{HW}\cite{STYEY}. A consequence of
this identification is the natural conjecture that the WDVV--like
equations (\ref{bw}) be equivalent to the higher rank version of the identity
(\ref{81}), namely \cite{STYEY}
\be
u={i\over 4\pi b_1}\left({\cal F}-\sum_i{a^i\over 2}a_i^D\right),
\l{ext}\ee
that satisfies the equation \cite{Noi2}
\be
{\cal L}_\beta u=u,
\l{hgdy}\ee
where  ${\cal L}_\beta$ is a second--order modular invariant operator.

Let us start by recalling the derivation of the WDVV--like equations
for the $SU(3)$ case whose curve and its extension to $SU(n)$ has been
derived in \cite{extension}.
Let us denote by $a^i= \langle \phi^i\rangle$ and $a_i^D=\langle 
\phi_i^D\rangle=\partial{\cal F}/\partial a^i$, $i=1,2$,
the vev's of the scalar component
of the chiral superfield and its dual. The effective couplings are given by
$\tau_{ij}=\partial^2 {\cal F}/\partial a^i\partial a^j$. We also
set $u^2\equiv u=\langle {\rm tr\, \phi^2}\rangle$, $u^3\equiv v=\langle{\rm tr
\, \phi^3}\rangle$ and $\partial_k\equiv \partial/\partial a^k$, 
$\partial_\alpha\equiv \partial/\partial u^\alpha$. 
The 
reduced Picard--Fuchs  equations (RPFE's) for $SU(3)$ are \cite{KLT}
\begin{equation}
{\cal L}_\beta \left(\begin{array}{c} a_i^D \\ a^i \end{array}\right)=0,
\qquad \beta=2,3,
\label{x1}\end{equation}
where
\begin{equation}
{\cal L}_2=(P/u)\partial_u^2+{\cal L},\qquad \qquad 
{\cal L}_3=(P/3)\partial_v^2+{\cal L},
\label{x2}\end{equation}
and $P=27(v^2-\Lambda^6)+4u^3$,
${\cal L}=12uv\partial_u\partial_v+3v\partial_v+1$.

Let us set 
$$
{\cal 
U}=u_2^2\partial_{11}-2u_1u_2\partial_{12}+u_1^2\partial_{22},\qquad
{\cal 
V}=v_2^2\partial_{11}-2v_1v_2\partial_{12}+v_1^2\partial_{22},
$$
$${\cal 
C}=(u_1v_2+v_1u_2)\partial_{12}-u_2v_2\partial_{11}-u_1v_1\partial_{22},
$$
and $D=u_1v_2-u_2v_1$,
where $\partial_{i_1...i_n}
\equiv {\partial^n/ \partial a^{i_1}...\partial a^{i_n}}$,
$u_i\equiv \partial_{i} u$ and
$v_i\equiv \partial_{i} v$.
 We have 
\begin{equation}
\left[12uv {\cal C}
+{1\over 3}P{\cal U}+D^2(1-a^i\partial_i)\right]{\cal F}_{l}=0
=\left[12uv {\cal C}+{1\over u} P
{\cal V}+D^2(1-a^i\partial_i)\right]{\cal F}_{l},
\label{x3}\end{equation}
where $l=1,2$ and ${\cal F}_{i_1...i_n}\equiv 
\partial_{i_1...i_n}{\cal F}$.
Subtracting the LHS from the RHS of Eqs.(\ref{x3}), we obtain
\begin{equation}
A_l\equiv x_{11}{\cal F}_{22l}+x_{22}{\cal F}_{11l}-2x_{12}
{\cal F}_{12l}=0, 
\label{x5}\end{equation}
where $l=1,2$ and
\be
x_{ij}=3v_iv_j-uu_iu_j.
\l{nnnnn}\ee
Next, considering 
$$
A_1\left(y_{22} {\cal F}_{112} -2y_{12} {\cal F}_{122}
+y_{11}{\cal F}_{222}\right)-A_2
\left(-2y_{12} {\cal F}_{112} +y_{11} {\cal F}_{122}
+y_{22}{\cal F}_{111}\right)=0,
$$
where $y_{jk}$ are arbitrary parameters, we get
\be
{\cal F}_{ikl}\eta^{lm}{\cal F}_{mnj}=
{\cal F}_{jkl}\eta^{lm}{\cal F}_{mni},
\label{wdvvv}\ee
for $i,j,k,n=1,2$, where
\be
\eta^{lm}=
\left(\begin{array}{c}2x_{22}y_{12}-2x_{12}y_{22}\\x_{11}y_{22}-
x_{22}y_{11}
\end{array}\begin{array}{cc}x_{11}y_{22}-x_{22}y_{11}
\\2x_{12}y_{11}-2x_{11}y_{12}\end{array}\right).
\label{hfgty}\ee
For any choice of the parameters 
$y_{jk}$, there is only one nontrivial equation in 
(\ref{wdvvv}) which can be rewritten in the form
\be
\eta^{11}\Theta_{11}+2\eta^{12}\Theta_{12}+\eta^{22}\Theta_{22}=0,
\l{oiuhd}\ee
where 
$\Theta_{ij}=\left(
{\cal F}_{11i}{\cal F}_{22j}+{\cal F}_{11j}{\cal F}_{22i}\right)/2-
{\cal F}_{12i}{\cal F}_{12j}$,
which satisfies the identity
\be
2{\cal F}_{12l}\Theta_{12}={\cal F}_{22l}\Theta_{11}
+{\cal F}_{11l}\Theta_{22},\qquad l=1,2.
\l{osita}\ee

The fact that $\tau_{ij}$ is dimensionless implies that
\be
(\ldl + \Delta_{u^\gamma})\tau_{ij}=0,
\l{oidx1}\ee
where 
\be
\Delta_{u^\gamma}= 
\sum_{\gamma=2}^n \gamma u^\gamma{\partial \over \partial u^\gamma},
\l{scalingop}\ee
is the scaling invariant vector field. 

Let us consider the $beta$--function
\be
\beta_{ij}=\Lambda {\partial \tau_{ij}\over\partial 
\Lambda}|_{u^\alpha,u^\gamma,\ldots}.
\l{iosq}\ee
By (\ref{oidx1}) we have for $SU(3)$
\be
\beta_{ij}=-  \left(2u{\partial a^k\over \partial u}+ 3v{\partial
a^k\over \partial v}\right){\cal F}_{ijk}.  
\l{iosqeddu}\ee
Let us denote by $\beta^{ij}$ the inverse of the matrix $\beta_{ij}$.
Setting
\be
\eta^{ij}=\beta^{ij},
\l{arizummolo}\ee
implies the equation \be
x_{22}\beta^{22}+x_{11}\beta^{11}+2x_{12}\beta^{12}=0.
\l{arizammolo}\ee
This equation, which arises as a the consistency condition, is equivalent 
to $A_1=0=A_2$ and
\be
{\cal F}_{ikl}\beta^{lm}{\cal F}_{mnj}=
{\cal F}_{jkl}\beta^{lm}{\cal F}_{mni},
\label{wdvvvxxxx}\ee
for $i,j,k,n=1,2$, is an identity.

We now consider the $\beta$--function for $SU(n)$ with $n\geq 4$.
We have
\be
\beta_{ij}=-\Delta^k{\cal F}_{kij},  
\l{iosqeddddu}\ee
where
\be
\Delta^k=\Delta_{u^\gamma}a^k= 
\sum_{\gamma=2}^n \gamma u^\gamma{\partial a^k\over \partial u^\gamma}, 
\qquad k=1,\ldots, n-1.
\l{scalingopcona}\ee
Let us set
\be
F_i=\F_{ijk}.
\l{oh}\ee
For the inverse of the beta matrix function $\beta_{ij}$ we have
\be
\beta^{ij}= -(\Delta^k F_{k})^{-1}_{ij},  
\l{iosqedddduc}\ee

We now show that Eq.(\ref{wdvvvxxxx}) holds also
for $SU(n)$ $n> 3$ (and is no longer an
identity). In particular, Eq.(\ref{wdvvvxxxx}) can be derived
by the WDVV equations in \cite{MAMIMO}. These have
the form
\be
F_i F_k^{-1} F_j= F_j F_k^{-1} F_i.
\l{fhef}\ee
These equations have the property of
being invariant if  
\be
F_k^{-1} \longrightarrow \beta^{-1}=-(\Delta^k F_{k})^{-1}. 
\label{wdvvvxxxxt}\ee
To see this, we simply observe that
by (\ref{fhef}) one has $F_i^{-1} F_k F_j^{-1}= 
F_j^{-1} F_k F_i^{-1}$,
so that $F_i^{-1} \beta F_j^{-1}= 
F_j^{-1} \beta F_i^{-1}$, implying
\be
F_i \beta^{-1} F_j= F_j \beta^{-1} F_i,
\l{dazzzozo}\ee
that is
\be
{\cal F}_{ikl}\beta^{lm} {\cal F}_{mnj}=
{\cal F}_{jkl}\beta^{lm}{\cal F}_{mni}.
\l{disarmingresult}\ee

\vspace{1cm}

\noindent
{\bf Acknowledgements}. It is a pleasure to thank
G. Bonelli, J. Isidro and M. Tonin for discussions.
MM was  supported in part by 
the European Commission TMR programme ERBFMRX--CT96--0045.

\end{document}